\documentstyle[12pt]{article}
\title{Phenomenology of effective gravity}
\author{G.E. Volovik\\
Low Temperature Laboratory,
Helsinki University of Technology\\
P.O.Box 2200, FIN-02015 HUT, Finland\\
and\\
L.D. Landau Institute for Theoretical Physics,
  Moscow\\
}
\begin{document}
\maketitle
\begin{abstract}
{ The  cosmological constant
is not an absolute constant. The gravitating
part of the vacuum energy is adjusted to the energy density of
matter and to other types of the
perturbations of the vacuum. We discuss how the vacuum energy
responds (i) to the curvature of space in the  Einstein closed
Universe; (ii) to the expansion rate in the de Sitter Universe;
and (iii) to the rotation in the G\"odel Universe. In all these
steady state Universes, the gravitating vacuum energy is zero in
the absence of the perturbation, and is proportional to the
energy density of perturbation. This is in a full agreement
with the thermodynamic Gibbs-Duhem relation applicable to any
quantum vacuum. It demonstrates that (i) the cosmological
constant is not huge, since according to the Gibbs-Duhem
relation the contribution of zero point fluctuations to the
vacuum energy is cancelled by the trans-Planckian degrees of
freedom; (ii)  the cosmological constant is non-zero, since the
perturbations of the vacuum state induce the non-zero vacuum
energy; and (iii) the gravitating vacuum energy is on the order
of the energy density of matter and/or of other perturbations.
We also consider the vacuum response to the non-steady-state
perturbations. In this case the Einstein equations are modified
to include the non-covariant corrections, which are responsible
for the relaxation of the cosmological constant. The connection
to the quintessence is demonstrated. The problem of the
energy-momentum tensor for the gravitational field is discussed
in terms of effective gravity. The difference between the
momentum and pseudomomentum of gravitational waves in general
relativity is similar to that for sound waves in hydrodynamics.
  }
\end{abstract}

\section{Introduction}

Observations demonstrate that the  pressure of the vacuum in our
Universe is very close to zero as
compared to the natural value which follows from the Planck
energy scale,
$P \ll  P_{\rm Planck}$. The natural value $P_{\rm
Planck}\sim \pm  E_{\rm Planck}^4$ for the vacuum pressure and
vacuum energy, obtained by summing the zero point energies of
quantum fields, is by about 120 orders of magnitude exceeds the
experimental limit. This is the main cosmological constant
problem
\cite{Weinberg2,Sahni,Peebles,Padmanabhan}.

Exactly the same `paradox' occurs in any quantum liquid
(or in any other condensed matter). The
experimental energy of the ground state of, say, the
quantum liquid at $P=0$ is zero. On the other
hand, if one starts calculating the vacuum energy
summing the energies of all the positive and negative
energy modes up to the natural cut-off energy scale, one obtains
the huge energy on the order of $E_{\rm Planck}^4$, where the
role of the Planck energy scale is played by the Debye
temperature. However, there is no real paradox in quantum
liquids, since if in addition to the sub-Planckian modes one
adds the trans-Planckian (microscopic, atomic) modes one
immediately obtains the zero value, irrespective of the details
of the microscopic physics
\cite{Book}. Thus the fully microscopic consideration restores
the Gibbs-Duhem relation, $\rho=-P$, between the  energy
(the relevant thermodynamic potential) and the pressure
of the quantum liquid at $T=0$. This Gibbs-Duhem relation ensures
the nullification of the energy of the vacuum state, $\rho=0$, if
the external pressure is zero.

This is the main message from condensed matter to the
physics of the quantum vacuum:  One should not worry
about the huge vacuum energy, the trans-Planckian
physics with its degrees of freedom will do all the
job of the cancellation of the vacuum energy without
any fine tuning and irrespective of the details of the
trans-Planckian physics.

There are
other messages which are also rather
general and do not depend much on details of the
trans-Planckian physics.  For example, the gravitating
energy of the perturbed vacuum is non-zero, and it is
proportional to the energies related to perturbations.
Thus, if the Planck ether (the vacuum) also obeys the
Gibbs-Duhem relation, then the
cosmological constant $\Lambda$ is not a constant but is an
evolving physical parameter, and our goal is to find its
response to different perturbations of the quantum vacuum. We
first consider the response of  $\Lambda$ to the
steady state perturbations, such as  the spatial curvature,
steady state expansion and rotation.  The response can be
obtained either using the Einstein equations or in a pure
phenomenological way. Then we consider the response to the
time-dependent perturbations of the vacuum. For that we need
some modification of the Einstein equations to allow
  $\Lambda$ to vary in time, and we shall discuss the
phenomenology of this modification. Ref. \cite{Evolution}
represents the short version of this paper.

\section{Gravity as perturbation of quantum vacuum}

\subsection{Einstein theory in standard formulation}

Let us start with the non-dissipative equations for
gravity -- the Einstein equations. They are obtained from
the action:
\begin{equation}
S=S_{\rm E}+S_\Lambda+S^{\rm M}~~'
\label{EinsteinAction1}
\end{equation}
where $S^{\rm M}$ is the matter action,
\begin{equation}
S_{\rm E}= -{1\over
16\pi G}\int d^4x\sqrt{-g}{\cal R}~~'
\label{EinsteinAction2}
\end{equation}
is the Einstein curvature action, and
\begin{equation}
  S_\Lambda=
-{\Lambda\over 8\pi G}\int d^4x\sqrt{-g}~,
\label{EinsteinAction3}
\end{equation}
where
$\Lambda$  is the cosmological
constant introduced by Einstein in 1917
\cite{EinsteinCosmCon}, which now got some experimental
evidences. The variation over the metric $g^{\mu\nu}$ gives the
Einstein equations
\begin{equation}
2\delta S=\delta g^{\mu\nu}\sqrt{-g} \left[
-{1\over 8\pi G}\left( R_{\mu\nu}-{1\over 2}{\cal
R}g_{\mu\nu}\right)+ {\Lambda \over 8\pi G} g_{\mu\nu}
+T^{\rm M}_{\mu\nu} \right]=0~,
\label{EinsteinEquation1}
\end{equation}
where $T^{\rm M}_{\mu\nu}$ is the
energy-momentum tensor for matter. The
Eq.(\ref{EinsteinEquation1}) has been originally
written in the form where the matter
term  is on the rhs of the equation, while the
gravitational terms which contain two constants $G$ and
$\Lambda$ are on the lhs (see  the Einstein
paper
\cite{EinsteinCosmCon}):
\begin{equation}
  {1\over 8\pi G}\left(G_{\mu\nu}  -
\Lambda g_{\mu\nu}\right)
  =T^{\rm M}_{\mu\nu}~,
\label{EinsteinEquation}
\end{equation}
where
\begin{equation}
  G_{\mu\nu}=
R_{\mu\nu}-{1\over 2}{\cal R}g_{\mu\nu}  ~
\label{EinsteinTensor}
\end{equation}
is the Einstein tensor. The form (\ref{EinsteinEquation}) of the
Einstein equation implies that the matter fields serve as the
source of the gravitational field.

\subsection{Cosmological constant as vacuum energy}

Later the cosmological constant term was moved to the
rhs of the Einstein equation:
\begin{equation}
  {1\over 8\pi
G}G_{\mu\nu}=T^{\rm
M}_{\mu\nu}+T^{\rm vac}_{\mu\nu}~,
\label{EinsteinEquation10}
\end{equation}
where in addition to the matter it became
another source of the gravitational field
  and got the meaning of
the energy-momentum tensor of the vacuum
\cite{Bronstein}:
\begin{equation}
  T^{\rm vac}_{\mu\nu}
=\rho^{\rm vac} g_{\mu\nu}~~,~~\rho^{\rm vac}={\Lambda\over
8\pi G}
  ~,
\label{VacuumEM}
\end{equation}
with $\rho^{\rm vac}$ being
the vacuum energy density and $P^{\rm vac}=-\rho^{\rm vac}$
being the vacuum pressure.

\subsection{Sakharov gravity as elasticity of quantum vacuum}

In the induced gravity introduced by
Sakharov \cite{Sakharov}, the gravity is the
elasticity of the vacuum. If it is the fermionic vacuum,
the effective action for the gravitational field is induced by
the vacuum fluctuations of the fermionic matter fields.
Such kind of the effective gravity emerges in the
fermionic quantum liquids of some universality class
\cite{Book}. In the induced gravity  the Einstein
tensor must be also moved to the matter side, i.e. to
the rhs, and the Einstein equations acquire the form
\begin{equation}
0=T^{\rm
M}_{\mu\nu}+T^{\rm vac}_{\mu\nu}+T^{\rm gr}_{\mu\nu}~.
\label{EinsteinEquation2}
\end{equation}
Here the tensor
\begin{equation}
  T^{\rm gr}_{\mu\nu}
=-{1\over  8\pi G}G_{\mu\nu}
  ~
\label{Curvature}
\end{equation}
has the meaning of the stress-energy tensor produced
by  deformations of the (fermionic) vacuum. It
describes such elastic deformations of the vacuum, which
distort the effective metric field $g_{\mu\nu}$ and
thus play the role of the gravitational field. As
distinct from the
$T^{\rm vac}_{\mu\nu}$ term which is of the 0-th order in
gradients of the metric, the $T^{\rm gr}_{\mu\nu}$
term is of the 2-nd order in gradients of
$g_{\mu\nu}$. The  higher-order gradient terms
also naturally appear in induced gravity.

\subsection{Conservation of energy and momentum}

In the induced gravity the free gravitational field is absent,
since there is no gravity in the absence of the quantum
vacuum (in the same manner as there is no sound waves in the
absense of the quantum liquid). Thus the total energy-momentum
tensor comes from the original (bare) fermionic degrees of
freedom. That is why all the three contributions to the
energy-momentum tensor are obtained by the variation of the
total fermionic action over
$g^{\mu\nu}$:
\begin{equation}
  T^{\rm total}_{\mu\nu}={2\over  \sqrt{-g}}{\delta
S\over\delta g^{\mu\nu}}= -{1\over 8\pi
G}G_{\mu\nu}  +
\rho^{\rm vac} g_{\mu\nu} +T^{\rm M}_{\mu\nu} ~.
\label{TotalEMT}
\end{equation}
According to the variational principle, $\delta
S/\delta g^{\mu\nu}=0$, the total
energy-momen\-tum tensor is zero, which gives rise to
the Einstein equation in the
form of Eq.(\ref{EinsteinEquation2}).

In the conventional description of general relativity, there
is a problem related to the energy-momentum tensor for
gravity: it is impossible to combine the general  covariance
with the conservation of energy and momentum. Attempts to
introduce the energy-momentum tensor for the gravitational
field led to various non-covariant objects called
pseudotensors (see e.g. \cite{Babak} and references
therein). In the induced gravity description,  there is no
contradiction between the covariance and the conservation
laws: each of three components (matter, gravity, and vacuum)
obey the covariant conservation law:
\begin{equation}
T^{\mu~{\rm
M}}_{\nu;\mu}=0~~,~~T^{\mu~{\rm
gr}}_{\nu;\mu}=0~~,~~T^{\mu~{\rm
vac}}_{\nu;\mu}=0~,
\label{Conservation laws1}
\end{equation}
while the total system
obeys both the covariant conservation law and the true
conservation law:
\begin{equation}
  \partial_\mu T^{\mu~{\rm total}}_{\nu}=0~.
\label{Conservation laws2}
\end{equation}
In general relativity, this equation is the consequence of the
fact that  $T^{\mu~{\rm
total}}_{\nu}=0$.

Actually, in induced gravity, the equation
$T^{\mu~{\rm total}}_{\nu}=0$ is valid only in the lowest orders
of the gradient expansion and can be violated by the
higher-order terms which come from the Planck-scale physics and
do not respect the general covariance. However, the
true conservation law, $\partial_\mu (T^{\mu~{\rm
covariant}}_{\nu}+T^{\mu~{\rm
noncovariant}}_{\nu})=0$, must be obeyed since it exists on the
microscopic level. This difference between the induced and
fundamental gravity can be used for the construction of the
post-Einstein equations describing the evolution of the induced
cosmological constant.

\subsection{Three components of `cosmic fluid'}

In induced gravity there is no much difference between
three contributions to the energy-momentum tensor:
all three components of the `cosmic fluid' (vacuum,
gravitational field, and matter) come from the original
bare fermions. However, in the low-energy corner, where the
gradient expansion for the effective action works, one can
distinguish between these contributions: (i) some part of the
energy-momentum tensor ($T^{\rm M}_{\mu\nu}$) comes from the
excited fermions -- quasiparticles -- which in the effective
theory form the matter. The other parts come from the fermions
forming the vacuum -- the Dirac  sea. The contribution from the
vacuum fermionscan be expanded in terms of the gradients of
$g_{\mu\nu}$-field. (ii) The zeroth-order term represents
the energy-momentum tensor of the homogeneous vacuum -- the
$\Lambda$-term. Of course, the whole Dirac sea cannot
be sensitive to the change of the effective
infrared fields $g_{\mu\nu}$: only small
infrared perturbations of the vacuum, which we are
interested in, are described by these effective fields.
(iii)  The higher-order terms describe the
inhomogeneous distortions of the vacuum state, which plays the
role of gravity. The second-order term in gradients, the stress
tensor $T^{\rm gr}_{\mu\nu}$, represents the curvature term in
the Einstein equations.

\subsection{Induced cosmological constant}

In
the traditional approach the cosmological constant is
fixed, and it serves as the source for the metric
field: in other words the input in the Einstein equations
is the cosmological constant, the output is  de Sitter
expansion, if matter is absent.
In effective gravity,
where the gravitational field, the matter fields, and
the cosmological `constant' emerge simultaneously in the
low-energy corner, one cannot say that one of these
fields is primary and serves as a source for the
other fields thus governing their behavor. The
cosmological constant, as one of the players,  adjusts
to the evolving matter and gravity in a self-regulating
way. In particular, in the absence of matter ($T^{\rm
M}_{\mu\nu}=0$) the non-distorted vacuum ($T^{\rm gr}_{\mu\nu}=0$) 
acquires zero cosmological constant,
since according to the `gravi-neutrality' condition
Eq.(\ref{EinsteinEquation2}) it follows from equations
$T^{\rm M}_{\mu\nu}=0$ and $T^{\rm gr}_{\mu\nu}=0$
that
$T^{\rm vac}_{\mu\nu}=0$. In this approach, the
input is the vacuum configuration (in a given example
there is no matter, and the vacuum is homogeneous), the
output is the vacuum energy. In contrast
to the traditional approach, here the gravitational
field and matter serve as a source of the induced
cosmological constant.

\subsection{Gibbs-Duhem relation and cosmological constant}

This conclusion is supported by the effective gravity
and effective QED which emerge in quantum liquids
or any other condensed matter system
of the special universality class \cite{Book}.
The nullification of the vacuum energy in the
equilibrium homogeneous vacuum state of the system is general
and does not depend on the microscopic atomic (trans-Planckian)
structure of the system. It  follows from the variational
principle, or more generally from the Gibbs-Duhem relation
applied to the equilibrium vacuum state of the system if it is
isolated from the environment.

The Gibbs-Duhem equation for the liquid
containing
$N$ identical particles relates the
energy
$E$ of the liquid and its pressure
$P$ in equilibrium:
\begin{equation}
E=TS+\mu N- PV~,
\label{GibbsDuhemRelation}
\end{equation}
where $T$, $S$, $\mu$ and $V$ are correspondingly the
temperature, entropy, chemical potential and volume of the
liquid. Let us consider the ground state (the quantum vacuum) of
the system, i.e. the state without quasiparticles or other
excitations of the quantum vacuum. This ground state also
represents the quantum vacuum for the effective quantum field
theory of the interacting Fermi and Bose quasiparticles, which
emerges in the liquid at low energy. The relevant Hamiltonian,
whose gradient expansion gives rise to the action for the
effective quantum field theory and effective gravity, is
${\cal H}-\mu {\cal N}$ \cite{AGDbook}.  Thus the
energy of the quantum vacuum is
\begin{equation}
  E^{\rm vac} =E-\mu
N=\langle {\rm vac}|{\cal H}-\mu {\cal N}
|{\rm vac} \rangle
~.
\label{RelevantVacuumEnergy}
\end{equation}
Then
from the Gibbs-Duhem relation (\ref{GibbsDuhemRelation})  at
$T=0$ one obtains the equation of state for the
equilibrium quantum vacuum:
\begin{equation}
\rho^{\rm vac} \equiv {E^{\rm vac}\over V} =- P^{\rm vac}~,
\label{GibbsDuhemRelation2}
\end{equation}
which is the same as for the quantum vacuum in general
relativity. For the isolated homogeneous
vacuum state, the external pressure is absent and one obtains
the nullification of the vacuum energy
\begin{equation}
\rho^{\rm vac}=-P^{\rm vac}=0
~.
\label{Nullification}
\end{equation}
  This means that
in the effective gravity emerging in quantum liquid one has
$T^{\rm vac}_{\mu\nu}=0$, if the vacuum is in complete
equilibrium. This is valid for any quantum liquid. If the liquid
contains
  several different species $a$ of atoms, fermionic or bosonic,
the corresponding  Gibbs-Duhem relation and the relevant vacuum
energy are
\begin{equation}
E=TS- PV+\sum_a \mu_a N_a~,~ \rho^{\rm vac} ={1\over V}\langle
{\rm vac}|{\cal H}-\sum_a\mu_a {\cal N}_a|{\rm vac} \rangle~.
\label{GibbsDuhemSeveralSpecies}
\end{equation}
They again lead at $T=0$ to the equation of state
(\ref{GibbsDuhemRelation2}) for the vacuum, and to
the nullification of the vacuum energy of an isolated system.

Since these thermodynamic arguments are valid for any quantum
liquid, one may expect that they are valid for the Planck ether
too, irrespective of the details of its structure. If so, then
extending this general rule to the quantum vacuum of our
Universe, one obtains that the non-perturbed vacuum is not
gravitating, i.e. in the quiescent flat Universe without matter
one has
$\Lambda=0$ without any fine tuning. This means that
the contributions to the vacuum energy from the sub-Planckian
and trans-Planckian degrees of freedom exactly cancel each other
due to the thermodynamic identity, while each contribution is
huge, being of the fourth order of the Planck energy scale:
\begin{equation}
  \rho^{\rm vac}_{\rm sub}\sim \rho^{\rm vac}_{\rm
trans}\sim \pm E_{\rm Planck}^4~~~,~~~ \rho^{\rm vac}_{\rm sub}+
\rho^{\rm vac}_{\rm trans}=0~.
\label{SubTrans}
\end{equation}

\subsection{Cosmological constant from vacuum perturbations}

The equation (\ref{Nullification}) is valid only at zero
temperature. If $T\neq 0$, the same Gibbs-Duhem relation
demonstrates, that the vacuum energy is proportional to the
thermal energy of the quantum liquid. In the case of massless
quasiparticles one obtains that $\rho^{\rm vac}\sim T^4$, i.e.
the induced vacuum energy density is proportional to the energy
density of matter (let us recall, that in quantum liquids the
role of matter is played by the quasiparticle excitations).

  Now let us consider how this occurs in general relativity, i.e.
how the vacuum energy (the cosmological constant) responds to
the perturbations of the vacuum, caused by matter, expansion,
spatial curvature and rotation. The Einstein equation does not
allow us to obtain the time dependence of the cosmological
constant, because of the Bianchi identities
$G^{\nu}_{\mu;\nu}=0$ and covariant conservation law for matter
fields (quasiparticles)
$T^{\nu{\rm M}}_{\mu;\nu}=0$ which together lead to
$\partial_\mu\Lambda=0$. But the general relativity allows us to
obtain the values of the cosmological constant in different
  stationary or steady state Universes, such as the Einstein
static closed  Universe
\cite{EinsteinCosmCon}, the de Sitter expanding Universe
\cite{deSitter}, and the G\"ode lrotating  Universe
\cite{Goedel}. In other words, we shall study the response of
the cosmological constant to curvature, expansion and rotation.

We discuss how the values of the cosmological constant can be
obtained without solving the Einstein equations.  This can be
done in a purely phenomenological way, in which the equilibrium
conditions and equations of state for the three components of
the system (vacuum, gravitational field and matter) are used.

\section{Robertson-Walker metric and its energy momentum tensor}

\subsection{Einstein action}

Let us start with the Robertson-Walker metric
\begin{equation}
ds^2=dt^2-a^2(t)\left({dr^2\over 1-kr^2}+ r^2d\Omega^2\right)~.
\label{RobertsonWalker}
\end{equation}
Its  Ricci tensor $R^{\mu}_\nu$ and Ricci scalar ${\cal R}$ are
\begin{equation}
R_0^0= -3{\partial_t^2a\over a}~ , ~R_j^i= -\delta^i_j
\left({\partial_t^2a\over a} + 2H^2 +
{2k\over a^2}\right)~,~ {\cal R}= -6 \left({\partial_t^2a\over
a} + H^2 + {k\over a^2}\right)~, 
\label{RicciScalar}
\end{equation}
where  $H=\partial_t a/a$ is the time-dependent Hubble's
parameter. After
integration by parts the action for the gravitational
field becomes quadratic in $\partial_t a$:
\begin{equation}
S^{\rm gr}=-{1\over 16\pi G}\int \sqrt{-g}{\cal R}=-{1\over
16\pi G}\int \sqrt{-g}\tilde{\cal R}\equiv -{6\over 16\pi
G} \int   a^3 \left(H^2-{2k\over a^2}\right)~.
\label{RicciScalar2}
\end{equation}
The Einstein tensor $G^{\mu}_{\nu}=
R^{\mu}_{\nu}-(1/2){\cal R}\delta^{\mu}_{\nu}$ is
\begin{equation}
G_0^0= 3 \left( H^2 + {k\over
a^2}\right)~,~ G_j^i=   \delta^i_j
\left(2{\partial_t^2a\over a} + H^2 +
{k\over a^2}\right)~.
\label{EinsteinTensor}
\end{equation}

\subsection{Energy-momentum tensor for gravitational field}

The corresponding stress-energy tensor of the
gravitational field,
$T_{\mu\nu}^{\rm gr}  =-G_{\mu\nu}/8\pi G$, can be
represented in terms of the energy density $\rho^{\rm gr}$ and
the partial pressure $P^{\rm gr}$ of this gravitational
component of the `cosmological fluid':
\begin{eqnarray}
T_{\mu\nu}^{\rm gr}=\rho^{\rm gr}u_\mu u_\nu+
P^{\rm gr}(u_\mu u_\nu-g_{\mu\nu})
\label{CurvatureEnergyMomentumTensor1}\\
= {1\over 8\pi G}\left(  -3
u_\mu u_\nu
\left(H^2 + {k\over a^2}\right)+
(u_\mu u_\nu-g_{\mu\nu})
\left(2{\partial_t^2a\over a} + H^2 +
{k\over a^2}\right)\right)~.
\label{CurvatureEnergyMomentumTensor2}
\end{eqnarray}
Here $u^\mu$ is the
4-velocity of the `cosmological fluid'; in
the comoving reference frame used it is $u^\mu=\delta^\mu_0$.
We shall show that for all three Universes considered
below, the energy-momentum tensor $T_{\mu\nu}^{\rm gr}$ makes
sense, so that $\rho^{\rm gr}$ and $P^{\rm gr}$ do play a role,
correspondingly, of the energy density stored in the
gravitational field, and the partial pressure thermodynamically
related to this energy.
Later in Sec. \ref{Gravitons} we shall discuss $T_{\mu\nu}^{\rm
gr}$ for gravitational waves.

\section{Einstein static Universe}

\subsection{Equation of state for gravitational field}

For a static Einstein Universe, ${\dot a}=0$, and from
Eq.({\ref{CurvatureEnergyMomentumTensor2}) one has
\begin{equation}
P^{\rm gr}=-{1\over 3}\rho^{\rm gr}= {k\over 8\pi G a^2}~.
\label{StaticUniverseEofS}
\end{equation}

The equations (\ref{StaticUniverseEofS}) can be obtained from
the pure phenomenology, i.e. without using and solving the
Einstein equations. The magnitude of the energy density stored
in the gravitational field component can be obtained directly
from the Einstein action, since in the static case the energy
density is the Lagrangian density with opposite sign:
\begin{equation}
\rho^{\rm gr}=-{\cal L}={1\over 16\pi G}
  {\cal R}=-{3k\over 8\pi G a^2}~.
\label{StaticUniverseGrEergy}
\end{equation}
Then the partial pressure of the gravitational
field is  obtained from the thermodynamic definition of pressure:
\begin{equation}
P^{\rm gr}= - {d(\rho^{\rm gr}a^3)\over d(a^3)}=-
{1\over 3}\rho^{\rm gr} ~.
\label{StaticUniverseEofS2}
\end{equation}
Let us stress that this equation of state is applicable only for
the gravitational field associated with the static
Robertson-Walker metric.

\subsection{Einstein solution from phenomenology}

The Einstein static solution can be obtained without solving the
Einstein equations by using the following
phenomenological equations:  (i) the  gravineutrality
condition, which states that the total gravitating energy
density vanishes,
\begin{equation}
\rho^{\rm gr}+\rho^{\rm vac}+
\rho^{\rm M}=0~;
\label{GravineutralityConditions}
\end{equation}
(ii) the equilibrium conditions which states that the pressure of
the system is zero,
\begin{equation}
P^{\rm gr}+P^{\rm vac}+
P^{\rm M}=0~;
\label{EquilibriumConditions}
\end{equation}
(iii) the equation of state for the vacuum component
\begin{equation}
P^{\rm vac}=-\rho^{\rm vac}=-{\Lambda\over 8\pi G} ~;
\label{9}
\end{equation}
(iv) the equation of state for the gravitational field component
(\ref{StaticUniverseEofS2}); and (v) the equation
(\ref{StaticUniverseGrEergy}) for the energy density stored by
the gravitational field. The latter two equations, (iv) and (v)
were also obtained from phenomenology.

From (i-v) it follows that for the general equation of state for
matter, the Einstein static Universe has the following
induced cosmological constant and the curvature as functions of
the matter fields:
\begin{eqnarray}
\Lambda_{\rm Einstein}= 8\pi G \rho^{\rm vac} = 4\pi G (\rho^{\rm 
M}+3P^{\rm M})~,
\label{LambdaGeneralEinstein1}\\
\rho^{\rm gr}=-{3k\over 8\pi G a^2}= -{3\over 2}(\rho^{\rm M}+ P^{\rm M})~.
\label{LambdaGeneralEinstein2}
\end{eqnarray}
The equation (\ref{LambdaGeneralEinstein2}) requires that,
unless the matter is unconventional, the parameter $k$ which
characterizes the curvature of the 3D space must be positive
($k=+1$), i.e. the static Universe with conventional matter is
closed.

\section{de Sitter solution  as a thermodynamic
equilibrium state}

We know that the static Einstein Universe is unstable. So let
us turn to expanding Universes. Can one apply the same
phenomenology of effective theory to the expanding or
inflating Universes? Let show that the phenomenology does work
in case of the dynamics of the flat Universe, i.e  with
$k=0$:
\begin{equation}
ds^2=dt^2-a^2(t)\left(dr^2+ r^2d\Omega^2\right)~.
\label{RobertsonWalker}
\end{equation}
If $k=0$ the Lagrangian (\ref{RicciScalar2}) is
quadratic in time derivative,
\begin{equation}
S^{\rm gr} =-{1\over
16\pi G}\int \sqrt{-g}\tilde{\cal R}\equiv -{6\over 16\pi
G} \int   a^3 {(\partial_t
a)^2\over a^2} ~.
\label{RicciScalar3}
\end{equation}
That is why the energy density
stored in the gravitational field equals the Lagrangian density:
\begin{equation}
\rho^{\rm gr}=  - {1\over 16\pi G}
\tilde {\cal R}=-{3\over 8\pi G}H^2~.
\label{ExpandingEenergyDensity}
\end{equation}
  Since $H$
is invariant under scale transformation $a\rightarrow \lambda a$,
the equation of state for the gravitational field component,
\begin{equation}
  P^{\rm gr}=-\rho^{\rm gr} ~,
\label{Expanding}
\end{equation}
is now
the same as for the vacuum component. The equations
(\ref{ExpandingEenergyDensity}) and (\ref{Expanding}), which we
obtained here from phenomenology, reproduce
equations (\ref{CurvatureEnergyMomentumTensor1})
and (\ref{CurvatureEnergyMomentumTensor2}) for the energy
momentum tensor of the gravitational field.

  The equilibrium state of the flat expanding Universe is now
obtained from the (i) gravineutrality condition
(\ref{GravineutralityConditions}), and (ii) equilibrium
condition  (\ref{EquilibriumConditions}), which together with
(iii) the equation of state for the vacuum component, $P^{\rm
vac}=-
\rho^{\rm vac}$, and (iv) the equation of state for the
  gravitational field component,
$P^{\rm gr}=  -\rho^{\rm gr}$, read
\begin{equation}
\rho^{\rm gr}+\rho^{\rm vac}+
\rho^{\rm M}=0~~,~~-\rho^{\rm gr}-\rho^{\rm vac}+P^{\rm M}=0~.
\label{EquilibriumConditions4}
\end{equation}
Since $P^{\rm M} \neq -\rho^{\rm M}$, the only solution
of the equations (\ref{EquilibriumConditions4}) is
\begin{equation}
\Lambda_{\rm de~Sitter}=
8\pi G \rho^{\rm vac} =-
8\pi G \rho^{\rm gr}=3H^2~~,~~P^{\rm M} =\rho^{\rm M}=0~.
\label{LambdaGeneralDeSitter}
\end{equation}
  This is de Sitter Universe without
matter, which is the steady state Universe
because the Hubble parameter $H$ is constant. In fact, the de
Sitter Universe can be described in terms of the stationary but
not static metric:
\begin{equation}
ds^2=dt^2- {1\over c^2}\left(d{\bf  r} - {\bf
v}dt\right)^2~,
\label{RobertsonWalkerFlatModified}
\end{equation}
where the frame dragging velocity
\begin{equation}
   {\bf
v}=H{\bf r}~.
\label{RobertsonWalkerFlatModified2}
\end{equation}

\section{Phenomenology of G\"odel Universe}

\subsection{Rotating Universe}

Let us now consider the different class of the Universes, the
rotating Universe. In principle, it is not necessary to know the
metric of this state, since it is enough to know, that it
represents the local rotation of the vacuum with angular
velocity $\Omega$. However, for completeness we present the
metric field obtained by G\"odel as the solution of Einstein
equations.
In cylindrical coordinate system \cite{Goedel}
\begin{eqnarray}
ds^2=  \left(dt +\Omega R^2  d\phi \right)^2
-{dR^2\over 1 +{R^2\over R_0^2}} - R^2\left(1 -{R^2\over
R_0^2}\right)d\phi^2 - dz^2
\label{3}
\\=   dt^2{ 1 +{R^2\over R_0^2}\over 1 -{R^2\over R_0^2}}-
{dR^2\over 1 +{R^2\over R_0^2}} - R^2 \left(1 -{R^2\over
R_0^2}\right)\left(d\phi +   {\Omega dt
\over  1 -{R^2\over R_0^2}}
\right)^2  - dz^2  ~,
\label{4}\\
R_0= {\sqrt{2}c\over \Omega}~.
\label{5}
\end{eqnarray}
When $R\ll R_0$, the metric corresponds to the Minkowski metric
in the frame rotating with angular velocity $\Omega$, that is
why the vacuum in this Universe is locally  rotating with
velocity  $\Omega$.
Thus the gravitational energy $\rho^{\rm gr}$ describes now the
  vacuum perturbation caused by rotation. In other words, the
energy stored in the rotation, which is the excess of the energy
with respect to the homogeneous vacuum state, represents the
energy of the gravitational field component in the G\"odel state.

As follows from the analogy with the vacuum state in condensed
matter, if the rotation velocity is small, the energy of rotation
can be written in general form as proportional to $\Omega^2$.
This means that the rotating vacuum must be characterized by the
angular momentum (spin) ${\bf S}$, which is proportional to
${\bf \Omega}$. The coefficient $\chi$ in the linear response of
the spin density ${\bf s}$ to the rotation velocity
\begin{equation}
{\bf s} =\chi {\bf \Omega}~,
\label{ReponseToRotation}
\end{equation}
plays the role of the spin susceptibility of the quantum vacuum.

  The energy stored in the gravitational subsytem is thus
\begin{equation}
\rho^{\rm gr} =    {1\over 2}\chi \Omega^2~.
\label{RotationalEnergy1}
\end{equation}

Note that the  energy density of the gravitational field, which
is stored in rotation, $\rho^{\rm gr}=(\chi/2)\Omega^2$, actually
represents the effect of a local rotation. The global
solid-body rotation corresponds to the Minkowski state in
the rotating frame. In such a state the curvature
${\cal R}=0$, and thus there is no energy associated with the
gravitational field. The curvature in the solid-body
rotating state is nullified due to the terms in ${\cal R}$, which
are the full space derivatives.  For the local rotations these
counter-terms are absent, since they disappear after the
integration by parts. This situation is similar to some
`paradoxes' related to the angular momentum in condensed matter.

Now we must find  the spin susceptibility of the vacuum, $\chi$,
which characterizes the local response of the vacuum spin
density to the rotation.

\subsection{Spin susceptibility
of the vacuum}

The spin
susceptibility of the vacuum is
\begin{equation}
  \chi=-{1\over 4\pi G\sqrt{-g}}  ~.
\label{chi}
\end{equation}
This diagravimagnetic response of the vacuum to rotation follows
from the gravitational interaction of two spins in the vacuum.
If two bodies at points ${\bf
r}$ and ${\bf r}'$ have spins ${\bf S}$ and ${\bf S}'$
respectively, their interaction
energy in the post-Newtonian approximation is
\cite{LandauLifshitz2}
\begin{eqnarray}
U^{\rm gr} = - {\bf S}\cdot{\bf \Omega}({\bf r})= - {\bf
S}'\cdot{\bf
\Omega}({\bf r'})= - {G\over \sqrt{-g}} ~ {{\bf S}\cdot {\bf S}'
- 3({\bf n}\cdot {\bf S})({\bf n}\cdot {\bf S}')\over |{\bf
r}-{\bf r}'|^3} ~,
\label{Interaction2Spins}
\\
\nonumber
{\bf n}={ {\bf r}-{\bf r}'\over  |{\bf
r}-{\bf r}'|}~,~ \sqrt{-g}={1\over c^3}~.
\end{eqnarray}
This interaction via the vacuum demonstrates that the  spin
susceptibility of the vacuum is given by  Eq.(\ref{chi}). The
sign of the gravitational interaction energy of spins in
Eq.(\ref{Interaction2Spins}) is opposite to that of the magnetic
dipole-dipole interaction of their magnetic moments. For
example, the magnetic dipole-dipole interaction of the electron
spins is
\begin{equation}
U^{\rm magn} =  {\hbar\alpha\over M^2\sqrt{-g}} ~ {{\bf S}\cdot
{\bf S}' - 3({\bf n}\cdot {\bf S})({\bf n}\cdot {\bf S}')\over
|{\bf r}-{\bf r}'|^3} ~,
\label{MagnInteraction2Spins}
\end{equation}
where $\alpha$ is the fine structure constant, and $M$ is the
rest energy of the electron.

The Eq.(\ref{chi}) can be also
obtained from the frame dragging effect caused by
the spinning body. In the post-Newtonian
approximation, the frame dragging velocity
${\bf v}$ (with $\nabla\times{\bf v} =2{\bf \Omega}$) caused by
a body at ${\bf r}=0$ with the spin ${\bf S}^0$ is
determined by the variation of the following energy:
\begin{equation}
{1\over 2}\chi \int \sqrt{-g} {\bf \Omega}^2 - {\bf
S}^0\cdot {\bf \Omega}({\bf r}=0) ~.
\label{FrameDraggingAction}
\end{equation}
The variation over ${\bf v}$ gives the following
frame dragging field around the body:
\begin{equation}
{\bf v} ={1\over 2\pi \chi \sqrt{-g}}  {{\bf S}^0\times {\bf
r}\over r^3} ~.
\label{FrameDraggingarounfBody}
\end{equation}
Comparing this with Eq.(9.5.18) of the book \cite{Weinberg}
\begin{equation}
{\bf v} =-2G {{\bf S}^0\times {\bf
r}\over r^3} ~,
\label{FrameDraggingarounfBody2}
\end{equation}
one obtains Eq.(\ref{chi}) for the spin susceptibility of the
vacuum.

\subsection{Equation of state for the gravitational field and
equilibrium state}

  Since the spin is conserved quantity, the relevant gravitational
energy which must be used to obtain the partial
pressure of the gravitational subsystem, must be
expressed in terms of the total spin ${\bf S}=V{\bf s}$, where
  $V$ is the volume of the system:
\begin{equation}
E^{\rm gr}=\rho^{\rm gr}V=    {{\bf S}^2\over 2\chi V}~.
\label{RotationalEnergy}
\end{equation}
The equation of state for the gravitational subsystem
is obtained from Eq.(\ref{RotationalEnergy}) by varying over
the volume at fixed total momentum ${\bf S}$:
\begin{equation}
P^{\rm gr}=-{dE^{\rm gr}\over dV}=  \rho^{\rm gr}~.
\label{EofSGoedel}
\end{equation}
Then from the phenomenological equations
(\ref{GravineutralityConditions}--\ref{9}) and (\ref{EofSGoedel})
one obtains the following response of the cosmological cosntant
to the matter field in the rotating Universe:
\begin{eqnarray}
\Lambda_{\rm Goedel}= 8\pi G \rho^{\rm vac} = -4\pi G
(\rho^{\rm M}-P^{\rm M})~,
\label{LambdaGeneralGoedel1}\\
\rho^{\rm gr}= -{1\over 2}(\rho^{\rm M}+ P^{\rm M})~.
\label{LambdaGeneralGoedel2}
\end{eqnarray}

For the cold matter, $P^{\rm M}=0$, the
induced cosmological constant is
\begin{equation}
  \Lambda_{\rm Goedel}= 8\pi G\rho^{\rm vac} =-  \Omega^2
=-4\pi G \rho^{\rm M} ~,
\label{StaticUniverseGoedel}
\end{equation}
where we used Eq.(\ref{RotationalEnergy1}): $\rho^{\rm gr}=\chi
  \Omega^2/2=- \Omega^2/8\pi G$. The equation
(\ref{StaticUniverseGoedel})  has been obtained by G\"odel from
the solution of the Einstein equations \cite{Goedel}.

\section{Modification of Einstein equation \\ and
relaxation of the vacuum energy}

\subsection{Cosmological constant as evolving parameter}

Till now we have not got any new information, since all these
well known results were originally obtained by solving the
Einstein equations. So, what is the point in the rederivation of
the old results?  This served to demonstrate that actually there
are no cosmological constant puzzles. The  cosmological constant
is not an absolute constant: we have seen that the gravitating
vacuum energy is adjusted to different types of the
perturbations of the vacuum in addition to the energy density of
matter: (i) to the curvature of space in the  Einstein closed
Universe; (ii) to the expansion rate in the de Sitter Universe;
and (iii) to the rotation in the G\"odel Universe. In all these
cases the gravitating vacuum energy is zero in the absence of
perturbations, being  proportional to the energy density of
perturbations. This is in a full agreement with the Gibbs-Duhem
relation applicable to any quantum vacuum, and shows that (i) the
cosmological constant is not huge, since according to the
Gibbs-Duhem relation the contribution of zero point
fluctuations to the vacuum energy is cancelled by the
trans-Planckian degrees of freedom; (ii)  the
cosmological constant is non-zero, since the perturbations of
the vacuum state induce the vacuum energy; and (iii) the
gravitating vacuum energy is on the order of the energy density
of matter and/or of other perturbations.

There are other consequences of this phenomenological approach.
For example, the Gibbs-Duhem
relation does not discriminate between  the false vacuum and
true vacuum. The only requirement is that  the vacuum state must
correspond to a local minimum or a saddle point of the
energy functional. This means that (iv) the false vacuum is also
non-gravitating if it is not perturbed. This leads to a rather
paradoxical conclusion: (v) if the cosmological pahse transition
from the false to true vacuum occurs at low temperature, the
cosmological constant is (almost) zero above the cosmological
phase transition, but below the transition  it will also become
zero after some transient period. This transient period can be
accompanied by the inflationary stage of the expanding Universe.

Since the cosmological constant
is not a constant, but an evolving parameter, which is ajusted
to the vacuum perturbations, the remaining cosmological constant
problem is (vi) to understand how it evolves in time.
The Einstein
equation does not allow us to obtain the time
dependence of the cosmological constant. This is because of the
Bianchi identities, $G^{\nu}_{\mu;\nu}=0$, and the covariant
conservation law for matter fields (quasiparticles),
$T^{\nu{\rm M}}_{\mu;\nu}=0$, which together lead to
$\partial_\mu\Lambda=0$. To describe the evolution of the
cosmological constant we must modify the Einstein equations to
allow $\Lambda$ to relax to its equilibrium value. Thus the
relaxation term must be added which violates the general
covariance. This correction comes from the trans-Planckian
physics, and thus it must contain the Planck energy scale. The
Planck physics can also violate the Lorentz invariance: the
dissipation implies the existence of the preferred reference
frame, which is the natural ingredient of the trans-Planckian
physics.

\subsection{Dissipation in Einstein equation}

The dissipation in the Einstein equation  can be introduced in
the same way as  in two-fluid
hydrodynamics \cite{Khalatnikov} which serves as the
non-relativistic analog of the self-consistent
treatment of the dynamics of the vacuum component (the superfluid
component of the liquid) and the matter component (the normal
component of the liquid) \cite{Book}. We must add the
dissipative part $T^{\rm diss}_{\mu\nu}$ to the total
energy-momentum tensor in Eq.(\ref{EinsteinEquation2}):
\begin{equation}
  T^{\rm
M}_{\mu\nu}+T^{\rm vac}_{\mu\nu}+T^{\rm gr}_{\mu\nu}+
T^{\rm diss}_{\mu\nu}=0~.
\label{EinsteinEquationDissipation}
\end{equation}
In contrast to the conventional dissipation of
the matter, such as viscosity and thermal conductivity, this
term is not the part of
$T^{\rm M}_{\mu\nu}$. It describes the dissipative
back reaction of the vacuum, which does not influence
the matter conservation law $T^{\nu{\rm
M}}_{\mu;\nu}=0$. The condensed-matter example of such
relaxation of the variables describing the fermionic
vacuum is provided by the dynamic equation for the
order parameter in superconductors -- the
time-dependent Ginzburg-Landau equation  which contains
the relaxation term (see e.g. the book
\cite{KopninBook}).

In the lowest order of the gradient expansion, the
dissipative part $T^{\rm diss}_{\mu\nu}$ of the stress
tensor describing the relaxation  of
$\Lambda$ must be proportional to the first time
derivative of $\Lambda$.
Since $T^{\rm diss}_{\mu\nu}$ is a tensor, the
general description of the vacuum relaxation requires
introduction of several relaxation times.  In the isotropic
space we have only two such relaxation parameters, in the energy
and pressure sectors:
\begin{equation}
T_{\mu\nu}^{\rm diss}=\left(\tau_1 u_\mu u_\nu+
\tau_2 (g_{\mu\nu}-u_\mu u_\nu)\right)\partial_t\Lambda~.
\label{DissMomentumTensor1}
\end{equation}
Here the 4-velocity $u_\mu$ selects the reference frame of the
trans-Planckian physics, which may or may not coincide with the
comoving reference frame.  In principle, the two reference
frames can be dynamically coupled, for example there can be a
mutual dissipative friction which forces the two frames to be
aligned in equilibrium. This would be analogous to the
Gorter-Mellink \cite{GorterMellink}  mutual friction force
between different components of a superfluid liquid in condensed
matter, and it could also serve as a source of dissipation.

The preferred reference frame is an important issue in
the effective gravity. As the condensed matter analogy suggests,
\cite{Book} there can be several reference frames: (i) the
reference frame of the absolute spacetime; (ii) the preferred
reference frame, in which the non-covariant corrections to the
effective action coming from the Planck-scale physics have the
most simple form; (iii) the comoving frame;  (iv) the frame of
the matter; and finally (v) the so-called frame of texture, which
corresponds to the preferred topological frame discussed in
\cite{Barrow}.

Further we assume that the system is close to equilibrium,
so that $u_\mu$ in Eq.(\ref{DissMomentumTensor1}) is the
4-velocity of the comoving reference frame. Then the next problem
is to find the relaxation parameters,
$\tau_1$ and
$\tau_2$, which can be the functions of the matter fields.
However, in principle, these functions can be treated as
phenomenological, which can be extracted from the observations.

Let us consider several the most simple examples of relaxation
of the cosmological constant.

\subsection{Cosmological constant as integration
constant}

Let us start with the most simple case when $\tau_1=\tau_2={\rm
constant} \equiv
\tau_\Lambda$. In this case the dissipative part of the stress
tensor is proportional to the metric tensor, $T_{\mu\nu}^{\rm
diss}=
\tau_\Lambda g_{\mu\nu} \partial_t\Lambda$.  The Bianchi
identities require that
$\partial_t(\Lambda +\tau_\Lambda \dot
\Lambda)=0$, which gives  $\Lambda +\tau_\Lambda
\partial_t\Lambda=\Lambda_0$. From equations
(\ref{DissMomentumTensor1}) and (\ref{VacuumEM})  it follows
that $T^{\rm vac}_{\mu\nu}+ T_{\mu\nu}^{\rm
diss}={\Lambda_0\over  8\pi G}g_{\mu\nu}$, i.e. the integration
constant
$\Lambda_0$ plays the role of the cosmological constant.
  The other examples when the
cosmological constant arises as an integration constant are
well known in the literature (see reviews
\cite{Weinberg2,Padmanabhan}).
In our example,
the vacuum  energy density is not a constant, but exponentially
relaxes to
$\Lambda_0$:
\begin{equation}
  \Lambda(t)=\Lambda_0 + \Lambda_1\exp\left(- \frac{t}{
\tau_\Lambda}\right)~,
\label{EvolutionLambda}
\end{equation}
where $\Lambda_1$ is another integration constant.
But experimentally we cannot resolve such evolution of the vacuum
energy, unless this evolution influences the equations for the
matter fields. That is why the dissipative part of the stress
tensor should not be proportional to the metric tensor, and we
really need $\tau_2\neq \tau_1$.

\subsection{Flat Universe with two relaxation parameters}

In such a general case  the dynamics of the cosmological
constant is determined by the
Einstein equations modified by the relaxation term
(\ref{DissMomentumTensor1}).  Let us consider this for a flat
Robertson-Walker Universe. For simplicity, we assume that the
Planck-scale reference frame is aligned with the
comoving frame. In this case one obtains the following
equations
\begin{equation}
- 3H^2+\Lambda +\tau_1 \dot \Lambda + 8\pi
G\rho^{\rm M}=0~,
\label{EinsteinTensorRW00M}
\end{equation}
\begin{equation}
  3H^2+2\dot H-\Lambda -\tau_2 \dot \Lambda+ 8\pi
Gp^{\rm M}=0~.
\label{EinsteinTensorRWijM}
\end{equation}
Because of the presence of the non-covariant relaxation term, the
covariant conservation law for matter does not follow now from
the Bianchi identities. That is why the above two equations
must be supplemented  by the covariant conservation law
to prevent the creation of matter:
\begin{equation}
a \frac{\partial}{ \partial a}(\rho^{\rm
M}a^3)=p^{\rm M}a^3~.
\label{CovariantConservation}
\end{equation}
Together with the equation of state for matter, these give four
equations for the four functions $H$, $\Lambda$, $\rho^{\rm M}$
and $P^{\rm M}$.

Let us discuss some consequences of these equations.

\subsection{Relaxation after cosmological phase
transition}

Let us start with the case when the
relaxation occurs only in the pressure
sector, i.e. $\tau_1=0$, and assume also that the
ordinary matter is cold, i.e. its pressure $p^{\rm
M}=0$, which gives
$\rho^{\rm M}\propto a^{-3}$. Then one finds two
classes of solutions: (i)
$\Lambda={\rm constant}$; and (ii)
$H=1/(3\tau_2)$. The first one corresponds to the
conventional expansion with the constant $\Lambda$-term and
the cold matter, so let us discuss the second solution,
$H=1/(3\tau_2)$.

In the case when
$\tau_2={\rm constant}$, one finds that the
$\Lambda$-term and the energy density of matter
$\rho^{\rm M}$ exponentially relaxe to $1/(3\tau_2^2)$ and
to 0 respectively:
\begin{equation}
H= \frac{1}{ 3\tau_2},~\Lambda(t)= \frac{1}{ 3\tau_2^2}
-  8\pi G\rho^{\rm M}(t), ~  \frac{\rho^{\rm M}(t)}{
\rho^{\rm M}(0)}= \exp\left(- \frac{t}{ \tau_2}\right).
\label{Evolution}
\end{equation}
Such solution describes the behavior after the
cosmological phase transition.

The cosmological phase transitions also impose the fine-tuning
problem for the cosmological constant.  If, for example, the
electroweak phase transition occurs according to the Standard
Model of the electorweak interactions (see e.g.
\cite{Langacker}), then after this phase transition the vacuum
energy is reduced by the value of the energy of the
Higgs field. This is about $10^{50}$ larger than the
observational upper limit for the vacuum energy. It means that
the primordial value of the vacuum energy (i.e. before the phase
transition) must be fine-tuned to cancel 50 decimal.

Let us consider  how the same problem is resolved in condensed
matter on the exampe of the phase transition  which
occurs at
$T=0$. The proper example is the first-order transition between
two superfluid vacua, A and B, in the $^3$He liquid at
$T=0$. According to the Gibbs-Duhem relation, which is
applicable to the false vacuum too, the relevant vacuum energy
before the transition, i.e. that of the false vacuum, is
$E^{\rm false~vac}=E-\mu^{\rm false~vac} N=0$.
Immedieately after the phase transition to the true vacuum, the
vacuum energy acquires the big negative value thus violating the
Gibbs-Duhem relation. However, after that the chemical potential
$\mu^{\rm false~vac}$ starts to relax to the new value $\mu^{\rm
true~vac}$ to restore the Gibbs-Duhem relation in a new
equilibrium. As a result, after some time the relevant energy of
the true vacuum also becomes zero:
$E^{\rm true~vac}=E-\mu^{\rm true~vac} N=0$.

Let us extend this scenario to the cosmological vacuum and the
cosmological phase transition. If this analogy is correct,  this
implies that if the temperature is low, the cosmological
`constant' $\Lambda$ is (almost) zero before the transition.
Immedieately after the transition it drops to the negative value;
but after some transient period it relaxes back to zero. The
equation (\ref{Evolution}) just corresponds to the latter stage.
But this solution demostrates that in its relaxation after the
phase transition, the
$\Lambda$-term  crosses zero and finally becomes
a small positive constant determined by
the relaxation parameter $1/\tau_2$ which governs the
exponential de Sitter expansion.

\subsection{Dark energy as dark matter}

Let us now allow  $\tau_2$ to vary. In condensed matter
the relaxation and dissipation are determined by
quasiparticles, which play the role of matter. According to
analogy, the relaxation term must be also determined by matter.
The inverse relaxation time must contain the Planck scale
$E_{\rm Planck}$ in the denominator, since the
relaxation of $\Lambda$ must disappear in the limit of infinite
Planck energy, when the general covariance is restored. The
lowest-order term,   which contains the
$E_{\rm Planck}$ in the denominator, is
$\hbar/\tau_2\sim
T^2/ E_{\rm Planck}$, where $T$ is the characteristic
temperature or energy of matter. In case of radiation it
can be written in terms of the radiation density:
\begin{equation}
   \frac{1}{  3\tau_2^2}=8\pi\alpha G\rho^{\rm M}~,
\label{MatterDependentTau}
\end{equation}
where $\alpha$ is the dimensionless parameter. If
Eq.(\ref{MatterDependentTau}) can be applied to the
cold baryonic matter too, then the solution of the
class (ii) becomes again
$H=1/(3 \tau_2)$, but now $\tau_2$ depends on the
matter field.  This solution gives the standard power
law for the expansion of the cold flat universe and the
relation between
$\Lambda$ and the baryonic matter $\rho^{\rm M}$:
\begin{equation}
a \propto           t^{2/3} ,~8\pi G \rho^{\rm
M}=  \frac{4}{
3 \alpha  t^2},~H= \frac{2}{
3t},~\Lambda=(\alpha-1)8\pi\ G\rho^{\rm M}.
\label{DarkEnergyMatterRelation}
\end{equation}
This solution is completely equivalent to the flat expanding
Universe without the cosmological constant, which follows from
the Einstein equations with the cold matter at critical density.
The reason is that the effective vacuum pressure in
Eq.(\ref{EinsteinTensorRWijM}), which comes from the vacuum and
the dissipative
$\Lambda$-terms, cancel each other,
$p_\Lambda= -(\Lambda+\tau_2\dot\Lambda)/8\pi G=0$. This means
that in this solution the vacuum behaves as the cold dark
matter. Altogether, the energy density of this cold dark matter
and that of the ordinary matter form the critical density
corresponding to the flat universe in the absence of the vacuum
energy:
\begin{equation}
\rho^{\rm vac}+\rho^{\rm
M} =\rho_c=   \frac{1}{
6\pi G t^2} ~.
\label{FlatColdUniverse}
\end{equation}
This demonstrates that in some cases the vacuum can
serve as the origin of the non-baryonic dark matter.

\subsection{Analog of quintessence}

These
examples are too simple to describe the real
evolution of the present universe and are
actually excluded by observations \cite{Sahni}. The
general consideration with two  relaxation functions
  is needed. In this general case, the effective equation of
state which comes from the reversible and
dissipative $\Lambda$-terms corresponds to the varying in time
parameter $w_Q$
\begin{equation}
w_Q(t)={ p_\Lambda \over  \rho_\Lambda} =-
{\Lambda+\tau_2\dot\Lambda\over \Lambda+\tau_1\dot\Lambda}~.
\label{Q}
\end{equation}
Such an equation of state is usually ascribed to
the quintessence, a kind of the scalar field, which provides
the varying negative pressure.  The recent observational bounds
on
$w_Q$ can be found,  for example, in Refs.
\cite{Pogosyan,Melchiorri}.

\section{On energy and momentum of gravitational waves}
\label{Gravitons}

Above we considered
three steady-state Universes and found that the
energy-momentum tensor of gravitational field
determined as
$T_{\mu\nu}^{\rm gr}$ in Eq.({\ref{Curvature}) makes
sense, so that
$\rho^{\rm gr}$ and $P^{\rm gr}$ do play a role,
correspondingly, of the energy density stored in the
gravitational field, and the partial pressure
thermodynamically related to this energy. Here we apply
this definition ({\ref{Curvature}) to the gravitational
waves.

In the absence of matter and cosmological cosntant the
Einstein equations read as $T_{\mu\nu}^{\rm gr}=0$,
which means that the gravitational
field in empty space has zero energy and momentum. At first
glance this leads to the paradoxical conclusion that the
gravitational wave does not carry energy and momentum.

The origin of this paradox is in the non-linear
nature of the Einstein equations, which encodes the self
interaction of the gravitational field in general relativity.
Let us consider the perturbation theory describing the
propagation of waves of small amplitude,
$g_{\mu\nu}=
\eta_{\mu\nu}+h_{\mu\nu}^{(1)}+h_{\mu\nu}^{(2)}$, where
$\eta_{\mu\nu}$ is the Minkowski flat metric and
$h_{\mu\nu}\ll \eta_{\mu\nu}$ is the perturbation.
In linear approximation, the equation $T_{\mu\nu}^{\rm
gr(1)}=0$ represents the wave equation for the function
$h_{\mu\nu}^{(1)}$ describing  gravitons with two
polarizations
$h_{12}^{(1)}$  and
$h_{22}^{(1)}=-h_{11}^{(1)}$
propagating, say, along the $z$-axis.

Now let us go to the second-order terms. There is the
contribution to $T_{\mu\nu}^{\rm gr(2)}$ which is quadriatic
in $h_{\mu\nu}^{(1)}$. It is non-zero and it represents the
energy and momentum of the graviton  (see
\cite{Weinberg} Sec. 10.3):
\begin{equation}
T^ {\rm graviton}_{\mu\nu}={k_\mu k_\nu\over 8\pi G}
~\left( |h_{12{\bf k}}]^2 + {1\over
4}|h_{22{\bf k}}-h_{11{\bf k}}]^2\right)~,
\label{EMGraviton}
\end{equation}
where $h_{{\bf k}}$ are the ampltudes of the propagating
waves. The equation $T_{\mu\nu}^{\rm gr(2)}=0$ in this
quadratic approximation is restored by the second-order
correction $h_{\mu\nu}^{(2)}$. It describes the response of
the gravitational field to the energy and momentum of the
graviton, which serve as a source of the additional
gravitational field (see
\cite{Weinberg} Sec. 7.6):
\begin{equation}
T^ {\rm graviton}_{\mu\nu}+ T^ {\rm
gravity~field}_{\mu\nu}=0~,~T^ {\rm
gravity~field}_{\mu\nu}=- \frac{1}{ 8\pi
G}G_{\mu\nu}^{(2)} ~.
\label{EMGraviton2}
\end{equation}
Here $G_{\mu\nu}^{(2)}$ is the Einstein tensor  of the
gravitational field $h_{\mu\nu}^{(2)}$ induced by the
gravitating graviton. It depends linearly on
$h_{\mu\nu}^{(2)}$:
\begin{eqnarray}
  2G_{\mu\nu}^{(2)}=
\partial^\alpha\partial_\alpha h_{\mu\nu}^{(2)}
- \partial^\alpha\partial_\mu h_{\nu\alpha}^{(2)}
- \partial^\alpha\partial_\nu h_{\mu\alpha}^{(2)}
+ \partial_\mu\partial_\nu h_{\alpha}^{\alpha(2)}
\nonumber
\\
- \eta_{\mu\nu}\left( \partial^\alpha\partial_\alpha
h_{\beta}^{\beta(2) }
-\partial^\alpha\partial^\beta h_{\alpha\beta}^{(2)}
\right)
.
\label{EMGraviton3}
\end{eqnarray}

  Thus, in spite of the
fact that the energy-momentum tensor of the gravitational
field is zero in empty space, $T^{{\rm
gr}}_{\mu \nu}=0$, it can be decomposed in
two terms which cancel each other:
the energy-momentum tensor of the graviton $T^ {\rm
graviton}_{\mu\nu}$ and that of the gravitational field
induced by the energy and momentum of the graviton, $T^ {\rm
gravity~field}_{\mu\nu}$. Of course, such separation is not
covariant, but any separation between different degrees of
freedom, i.e. between different subsystems, does not
necessarily obey the symmetry of the whole system. The mere
presence of a graviton violates the general covariance,
since it introduces the distinguished reference frame. The
graviton plays the same role as any other matter which gives
rise to the preferred reference frame comoving with
matter. However, the energy-momentum tensor of the whole
system, $T^{{\rm
gr}}_{\mu \nu}$, obeys both the true conservation law and
covariant conservation law, simply because it is zero, and
thus there is no contradiction between the general
covariance and the conservation of energy and momentum.

There is a deep difference between the covariant
energy-momentum tensor
  $T^{{\rm gr}}_{\mu \nu}$ and the non-covariant
energy-momentum tensor of the graviton $T^ {\rm
graviton}_{\mu\nu}$. The first one is obtained as the
functional derivative of the action with respect to the
  the whole metric, while the second one is
obtained by variation over the metric which serves as the
background metric for graviton, i.e. from which some part is
excluded -- the
oscillating part involved in the graviton.

A similar situation occurs in hydrodynamics, which also
represents the non-linear theory. The propagation of sound
wave in liquids is governed by the dynamical acoustic metric
provided by the density $\rho$ and velocity
${\bf v}$ fields of the moving liquid. These fields play the
role of the gravity field in general relativity, while the
sound waves -- propagating perturbations $\rho^{(1)}$ and
${\bf v}^{(1)}$ -- play the role of
gravitational waves
\cite{Stone2002,Book}.  Here again from
the  same initial field, now the  hydrodynamic fields $\rho$
and ${\bf v}$, two subsystems are formed: (i) the phonon and
(ii) the smooth acoustic metric, which governs the phonon
propagation and is influenced by the phonon due to back
reaction. Thus there are also two metrics at our disposal
when we discuss the momentum of the sound wave: the metric of
the fundamental spacetime in which the liquid lives, and the
acoustic metric of the effective spacetime in which the
phonon lives
\cite{Stone2002}. The `real' energy and momentum are obtained
by differentiating with respect to the `real'
(fundamental) metric, while those obtained by differentiation
with respect to the acoustic (effective) metric are called
the pseudoenergy and pseudomomentum; many paradoxes in
condensed matter physics arise if these two notions
are not discriminated (see references in \cite{Stone2002}).

The `real' momentum density of the liquid
associated with the sound wave (phonon)  in Eq.(13.52) of
\cite{Stone2002} consists of two parts:
\begin{equation}
{\bf p}^{\rm liquid} =\langle\rho^{(1)}{\bf
v}^{(1)}\rangle + {\bf v}\langle\rho^{(2)}\rangle ~.
\label{PhononMomentum}
\end{equation}
The first term in Eq.(\ref{PhononMomentum}) is the
pseudomomentum of the phonon (the analog of graviton), while
the second term represents the momentum coming from the back
reaction of the liquid (analog of the gravitational field
produced by the gravitating graviton). Here
$\langle\rho^{(2)}\rangle$ is the second-order perturbation
of the density of the liquid caused by the phonon, which is
the analog of
$h_{\mu\nu}^{(2)}$.

Because of the momentum conservation law, which comes from
the translational symmetry of the whole system, the total
momentum of the liquid must be always zero, ${\bf p}^{\rm
liquid}=0$. However, each of the two momenta, of the phonon
and of the liquid, is non-zero and can be measured
separately. The mere presence of a phonon in the liquid
violates the translational invariance, and thus the
pseudomomentum of the phonon is non-zero.

Condensed matter systems provide many examples, when the
energy-momentum tensor, though is well defined
on the microscopic level, cannot be localized when it
is written in terms of the limited number of the
infrared collective variables (see e.g. \cite{Ferromagnets}
for ferromagnets). One can push further the analogy between
the quantum vacuum in general relativity and the quantum
vacuum in condensed matter. One can imagine that the metric
$g_{\mu\nu}$ arises only in the low-energy corner while the
whole vacuum lives in the fundamental spacetime, Lorentzian
or Galilean, appropriate for the microscopic
(trans-Planckian) physics
\cite{Book}.  If it is so, then even the  covariant
energy-momentum tensor $T^{{\rm gr}}_{\mu \nu}$, which is
expressed in terms of the collective field $g_{\mu\nu}$, does
not describe the real energy and momentum of the
gravitational field. The latter remain unknown and will be
revealed only at high energy, where the non-covariant
corrections caused by the Planck physics will indicate the
preferred reference frame of the quantum vacuum.

\section{Discussion}

If the gravity does belong to the class of the emergent
phenomena \cite{Laughlin}, we can obtain some useful
consequences of that. The main message is that in the effective
gravity the equilibrium time-independent vacuum state without
matter is non-gravitating, i.e. its relevant vacuum energy, which
is responsible for gravity, is zero. On the other hand, if the
vacuum is perturbed,  the cosmological constant is
non-zero, and it is ajusted to the perturbations. If the case of
the steady state perturbations, the response of the cosmological
constant can be found from the Einstein equations. It appears,
however, that in some of these cases it is not necessary to solve
Einstein equations, to obtain this response. The response to the
curvature, steady state expansion and rotation can be obtained
from the purely phenomenological approach, as we discussed in
this paper. The response of the vacuum energy to matter in the
world without gravity (i.e. when the Newton constant $G=0$) was
discussed in \cite{UniverseWithoutGravity}.

If the general case of the time-dependent perturbations, the
cosmological constant is an evolving parameter rather than the
constant. The process of relaxation of the cosmological
constant, when the vacuum is disturbed and out of the thermal
equilibrium, requires some modification of the Einstein
equation, since the Bianchi identities must be violated to allow
the cosmological constant to vary.   In contrast to the
phenomenon of nullification  of the cosmological constant in the
equilibrium vacuum, which is the general property of any quantum
vacuum and does not depend on its structure and on details of the
trans-Planckian physics, the deviations from the general
relativity can occur in many different ways, since  there are
many routes from the low-energy effective theory to the
high-energy `microscopic' theory of the quantum vacuum. However,
it seems reasonable that such modification can be written in the
general phenomenological way, as for example the dissipative
terms are introduced in the hydrodynamic theory. Here we
suggested to describe the evolution of the $\Lambda$-term by two
phenomenological parameters (or functions) -- the relaxation
times, and demonstrated that this term ic responsible for the
quintessence.

This work was supported by
ESF COSLAB Programme and by the Russian Foundations for
Fundamental Research.

\end{document}